# On the evaluation of the suitability of the materials used to 3D print holographic acoustic lenses to correct transcranial focused ultrasound aberrations

**Marcelino Ferri[1*], José M. Bravo[1], Javier Redondo[2], Juan V. Sánchez-Pérez[1], Noé Jiménez[2], Sergio Jimenez-Gambín[2], Francisco Camarena[2]**

1) Centro de Tecnologías Físicas. Universidad Politécnica de Valencia. Camino de Vera S/N. 46020. Valencia. Spain
2) Instituto para la Gestión Integrada de las zonas Costeras. Universidad Politécnica de Valencia. Carretera Nazaret-Oliva S/N. 46730. Grao de Gandia. Valencia. Spain
* Correspondence: mferri@fis.upv.es   0034 963877000 (75245)

**Abstract:** The correction of transcranial focused ultrasound aberrations is a relevant topic for enhancing various non-invasive medical treatments. Nowadays, the most widely accepted method to improve focusing is the emission through multi-element phased arrays; however, a new disruptive technology, based on 3D printed holographic acoustic lenses, has recently been proposed overcoming the spatial limitations of phased arrays due to the submillimetric precision of the latest generation of 3D printers. This works aims to optimize this recent solution; particularly, the preferred acoustic properties of the polymers used for printing the lens are systematically analyzed, paying special attention to the effect of p-wave speed and its relationship to the achievable voxel size of 3D printers. Results from simulations and experiments clearly show that there are optimal ranges for lens thickness and p-wave speed, fairly independent of the emitted frequency, the transducer aperture, or the transducer-target distance, given a particular voxel size.

## 1. Introduction

Holography is a technique to reconstruct wave fields from the previous recording of their complete amplitude and phase information. Optical holograms have been widely developed in many scientific, technological and even artistic applications; however acoustic holograms have been notably less applied to interest cases. Most acoustic holographic techniques are based on phased arrays with a large number of transducers [1, 2], requiring sophisticated electronic circuits highly sensitive to calibration inaccuracies. However, new advances in metamaterials and the design capabilities of the state of art 3D printers allow the development of passive structures capable of shaping complicated acoustic fields [3-9].

An ambit in which precise holographic reconstruction of acoustic fields is gaining importance in recent years is in focused ultrasound for medical applications (FUS), mainly in non-invasive treatments where the ultrasound propagates through tissues with very different acoustic impedances, as is the case of transcranial propagation [8-16]. The first successful transcranial ablation of animal brain tissue was achieved by Fry and Goss in 1980 [16], since then, many clinical procedures have been improved by applying transcranial focused ultrasound. Treatment of brain tumours [17, 18], Alzheimer's disease [19] or Parkinson's disease [20] can be enhanced by this technique. The administration of FUS allows a local transient opening of the blood-brain barrier (BBB) [13-21] improving the delivery of pharmacological substances such as anticancer therapeutic drugs [22], neurotrophic factors [23], adeno-associated viruses [24, 25], and neural stem cells [12].

The success of these medical treatments is linked to an adequate reduction of skull-induced aberrations to achieve a precise ultrasound focusing. Early approaches included ultrasonic therapy after craniotomy [26] or less aggressive techniques such as the application of FUS without correction of aberrations, but radiating from regular and flat areas of the skull [27]. More recent approaches mainly rely on actively shaping the wavefront to correct bone-induced aberrations through the use of holographic lenses or multi-element transducer arrays whose phase can be adjusted individually. The required phase registration is performed by inverse propagation, measured or simulated, from target to transducer. Initially, phase patterns were assessed by physically placing a reference transducer inside the brain at target location [28, 29]; nowadays, the technique is absolutely non-invasive as acceptable phase patterns can be obtained by numerical simulation [14, 30-31].

The research aiming to improve the technique can be based both on (i) a better recording of the phase -associated with the knowledge of the physical variables of each element of volume scanned by medical imaging and with the numerical methods applied- and on (ii) a better reconstruction of the recorded field. In this work we are going to address the second issue. In fact, the aberration correction by multi-element phased arrays is subjected to technological limitations associated with the size and number of singular elements that can be implemented in the phased arrays. These limitations can be overcome by shaping the wavefront using high transparency 3D printed refractive ultrasound lenses [6, 8]. In fact, the work published by Maimbourg et al. [6] demonstrates that the spatial resolution, associated with the voxel size achieved by the latest generation of 3D printers, improves dramatically the space resolution of phased arrays, leading to better focusing. Therefore, in addition to the actual race towards creating arrays with an ever-increasing number of elements, the holographic acoustic lens approach emerge as a promising technology allowing submillimetric phase correction with limited cost.

In this work, assuming the relevance that 3D holographic lenses can take for transcranial FUS treatments in the near future, we aim to systematically evaluate the physical properties that the polymers used in their elaboration must have in order to reconstruct the desired acoustic field in an optimal way. There are trivial aspects, as that the lens must have maximum transparency, however, other aspects, such as those relating to the shape of the lens and the sound speed in the polymer-intimately linked to each other- require a more detailed quantitative study. In fact, in previous works, polymers with lower [6] or higher [8,9] than water wave speeds have been indistinctly used to build the lenses; without explicitly expressing a suitability criterion. In both slow and fast materials their p-wave speeds condition the thickness and shape of the lenses, and there are important aspects of lens quality associated with shape such as (i) the validity of the theoretical approach that assumes the hypothesis of ideally thin lenses, or (ii) the effect of voxel size on the printed lens, since the relative discretization of the lens is more abrupt in thin lenses. In order to give a quantitative response, or at least a guide to solve these questions, a series of numerical simulations and a validation experiment are proposed, the results of which are discussed and justified.

## 2. Materials and Methods

### 2.1. Physical parameters obtained from computerized tomographies

Numerical simulations were conducted using the data acquired by computerized tomography (CT) of a human head (freely available at repository cancerimagingarchive.net for scientific purposes) with an interslice spacing of 0.63 mm and an in-plane spatial resolution of 0.49 mm. 3D linear interpolation of the radiodensity known at each node of this parallelepiped mesh is performed to obtain a denser cubic simulation grid with a spatial step, $h = 0.244$ mm. Different excitation frequencies are applied in simulations, being the highest 760 kHz, so a ratio $\lambda/h > 8$ at water is ensured, which introduces acceptable numerical phase error in the simulations[32].

The elastodynamic damped equations for isotropic solids are applied at the whole computational domain, since the dynamics of the wave propagation in fluids can be defined as a particular case of solid with null shear modulus $G$. A set of four independent parameters must be known at each computational node, being these (i) the density in equilibrium $\rho$, (ii) the p-wave speed $c$, (iii) the attenuation coefficient $\alpha$, and (iv) the Poisson's ratio $\nu$. The two first parameters are obtained by means of the linear interpolations [6]

$$c(x,y,z) = c_{water} + \left(c_{bone} - c_{water}\right)\frac{HU(x,y,z) - HU_{\min}}{HU_{\max} - HU_{\min}} \tag{1}$$

$$\rho(x,y,z) = \rho_{water} + \left(\rho_{bone} - \rho_{water}\right)\frac{HU(x,y,z) - HU_{\min}}{HU_{\max} - HU_{\min}} \tag{2}$$

where HU is the radiodensity in Hounsfield units, $\rho_{water}$ and $c_{water}$ are respectively the density and p-wave speed of water at 21ºC, and $\rho_{bone}$ and $c_{bone}$ are respectively the density and p-wave speed of cortical bone at that temperature. Radiodensity values out of the interval 0 to 2400 HU were set respectively to these saturation values expected for water and cortical bone [33]. In accordance with bibliography [6, 34, 35, 36] the values considered for water and cortical bone are

$$c_{bone} = 3100\,m/s\,,\ \ \rho_{bone} = 1900\,kg/m^3\,,\ \ c_{water} = 1485\,m/s\,,\ \ \rho_{water} = 10^3\,kg/m^3\,.$$

Since the dependence of radiodensity on attenuation coefficient and Poisson's ratio has not been investigated in depth, we have applied a simplified approach for these parameters, obtaining an average value at each domain. For the attenuation coefficients at 760 kHz we have taken $\alpha_{brain}$=2 Np/m, and $\alpha_{skull}$= 60 Np/m. [8, 34, 37-38], and Poisson's ratio at the skull has been defined as uniform with a value $\nu_{skull}$ = 0.316 to achieve a constant relation between p-wave and s-wave speeds, $c_P/c_S$=27/14, as proposed by Hughes et al. [39].

### 2.2. Governing equations and numerical model

A linear centered elastodynamic FDTD method has been developed by the authors in Matlab (The MathWorks, Inc. Massachusetts. USA). The governing equations for both fluid and isotropic solid media, are [8]:

$$\frac{\partial \tau_{ij}}{\partial t} = (M - 2G)\delta_{ij}\left(\nabla \vec{u}\right) + G\left(\frac{\partial u_i}{\partial x_j} + \frac{\partial u_j}{\partial x_i}\right) \tag{3}$$

$$\rho\frac{\partial u_i}{\partial t} + \sigma \cdot u_i = \sum_j \frac{\partial \tau_{ij}}{\partial x_j} \tag{4}$$

where $u$ is the particle velocity, $\tau_{ij}$ are the components of the stress tensor, $\delta_{ij}$ is the Kronecker delta, $\rho$ is the density in equilibrium, $M$ and $G$ are respectively the p-wave and the shear moduli, and $\sigma$ is an artificial absorption parameter generating an exponential space dependent attenuation in isotropic media. The value of $\sigma$ is implemented to obtain the proper attenuation coefficient, for both solid and fluid media, attending to the relation proposed by Ferri et al [8]

$$\sigma = \rho \sqrt{\left(\omega + \frac{2c^2\alpha^2}{\omega}\right)^2 - \omega^2} \,, \tag{5}$$

where $\omega$ is the angular frequency, c the p-wave speed and $\alpha$, in Nepers per meter, is the absorption at the media. Shear modulus, G, is obtained from its known relation with Poisson's rate and p-wave modulus at solid media, and forced to zero in fluid domain.

Equations 3 and 4 are valid for both solid and fluid media. At liquid media, the null value of the shear modulus $G$ leads to a null value of the tangential stresses and to an equal value of the three axial stresses at each point, equivalent to the fluid pressure with opposite sign. The parameter $M$ represents respectively the p-wave modulus at solid domain and the bulk modulus at fluid; and it is obtained at any point as

$$M = \rho c^2 \tag{6}$$

The excitation signal, for both time forward and time reversal simulations, consists of a harmonic pressure excitation enveloped by a half Hanning window during the first $n$ cycles, with $n$ = 8. Excitation pressure is implemented in complex form, in order to facilitate the phase computation at the holographic registration surface, as

$$\tau_{11} = \tau_{22} = \tau_{33} = -p_o \sin^2\left(\frac{\omega}{4n}\min\left(t, \frac{2n\pi}{\omega}\right)\right)\left(\cos(\omega t) + j\sin(\omega t)\right). \tag{7}$$

The time step $\Delta t$ is adjusted to a 0.075 Courant–Friedrichs–Lewy (CFL) condition at water that ensures stability at the fastest evaluated holographic lens that is four times faster than water being also faster than cortical bone. The definition of a CFL linked to the water, instead of dependent on the fastest media as usual, attends to the fact that the numerical isotropy of the wave speed is CFL dependent [40]. Its oversized value intends to avoid favouring propagation in some lenses over others, as it is far from the optimum in all cases.

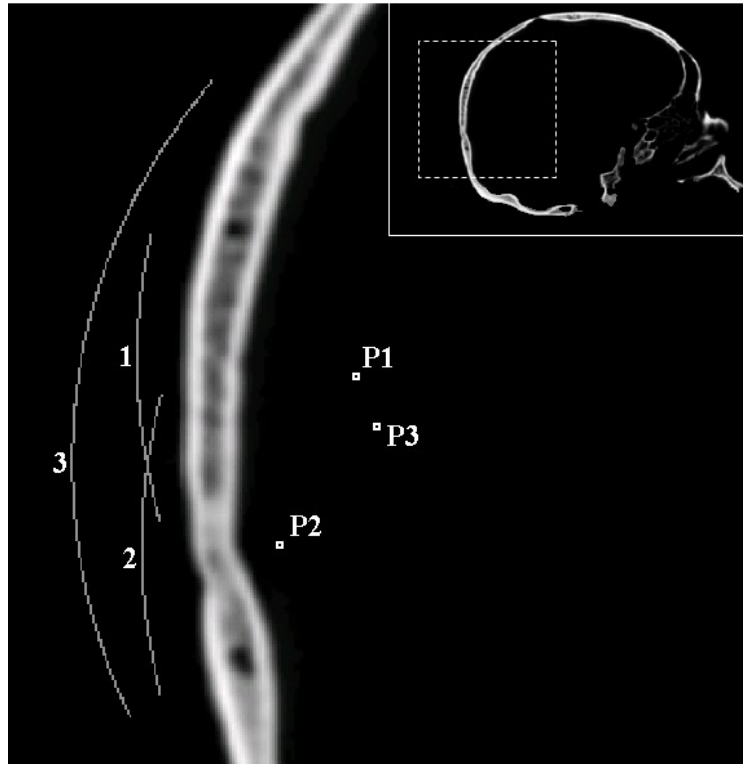

**Figure 1**. Schematic sagittal cross section of the skull, representing the positions of target points and registration surfaces considered in numerical simulations.

## 2.3. Holographic phase pattern registration

The lens shape is derived from the phase pattern obtained by simulation at the holographic registration surface, concentric to the surface where the transducer must be placed, at a distance equivalent to the nominal lens thickness (Figure 1). Properly, the holographic registration surface is a Cartesian discretized spherical or plane surface where the phase is obtained from the value of the pressure in complex form known at the Cartesian nodes of the computational grid. This pressure field is simulated by the time reversal propagation of a single frequency wave emitted at the target point. At the stationary state, achieved several periods after the wave reaches the registration surface, the numerically computed phase shift between any two points of the computational domain remains remarkably constant. Phases calculated at the holographic registration surface are not subjected to a process of unwrapping, as recommended in previous works [6, 8], since this work is focused in evaluating holographic Fresnel lenses capable of a maximum phase shift of $2\pi$. Instead, we just add an arbitrary value to the phase pattern to reduce the region of the lens affected by an abrupt $2\pi$ phase shift

## 2.4. Numerical design of holographic lens

The thickness at each point of the lens defined in spherical coordinates, $h(\theta,\varphi)$ is obtained by linear interpolation from the phase $\beta(\theta,\varphi)$ computed at the holographic registration surface, as follows

$$h(\theta,\varphi) = \frac{\beta(\theta,\varphi)}{2\pi}d \qquad \text{or} \qquad h(\theta,\varphi) = \frac{2\pi - \beta(\theta,\varphi)}{2\pi}d \qquad (8)$$

were the first expression is used for lenses made with materials with wave speed smaller than water and the second for fast lenses, and where $d$ is the nominal lens thickness obtained for both fast and slow lenses, as a function of its relative speed ($c_{ref} = c_{p\_lens}/c_{water}$), as

$$d = \left| \frac{c_{rel}}{c_{rel}-1} \right| \lambda_{water} \qquad (9)$$

where, $\lambda_{water}$ is the wavelength of the surrounding media (generally water) and the relative speed is obtained from the p-wave speed of the lens. The phases, $\beta(\theta,\varphi)$, required to obtain the thickness pattern of the lens, $h(\theta,\varphi)$, are computed after interpolating the phases, $\beta(x,y,z)$, at the Cartesian holographic registration surface to the closest points of a mesh defined in spherical coordinates centered at the transducer focus. If 3D lenses are implemented in simulations, their shapes are linearly interpolated back to Cartesian domain

Linear interpolation proposed in equation 8 is accurate if no inner reflections in both surfaces of the lens are considered, i.e if the acoustic impedance of the lens is reasonably similar to the acoustic impedance of the surrounding media (water). If this hypothesis of total transmission is not acceptable, the thickness of the lens should be obtained applying the Fabry-Pérot approach, as proposed by Jimenez-Gambín et al [9]. This approach attends to an implicit expression that should be solved numerically for each value of $\beta(\theta,\varphi)$. The relevance of the systematic error assumed by applying the explicit linear interpolation is discussed in later sections.

## 2.5. Experimental equipment and materials

The acoustical lens employed at the validation experiment was manufactured by stereolithographic 3D-printing techniques (Forms 2, Formlabs Inc., USA) using a photopolymer resin (Grey Standard Form 2, Formslabs Inc., USA) with a resolution of 100 µm, as shown in Figure 2(a). The material was post-cured after 3D printing process with a 1.25 mW/cm² of 405 nm LED light for 30 minutes at 60 °C. The acoustical properties of the material were obtained experimentally using a pulse-echo technique in a test cylinder with a height of 30 mm and radius 25 mm, resulting in a measured sound speed of $c_p$ = 2440.7 m/s and a density of $\rho$ =1162.0 kg/m³; as listed in Table 1 where data of a set of 3D printable polymers are presented [41-45].

The rest of devices applied consist of a single-element piezoelectric transducer (PZT26 Ferroperm Piezoceramics, Denmark), a signal generator (PXI5412, National Instruments, USA) amplified by a linear RF amplifier (1040L, ENI, Rochester, NY), a needle hydrophone (HNR-500, Onda) calibrated from 1 MHz to 20 MHz, a digitizer (PXI5620, National Instruments, USA) and a precision 3D micro-positioning system (OWIS GmbH, Germany) All the signal generation and acquisition processes were based on a NI8176 National Instruments PXI-Technology controller, which also controlled the micro-positioning system.

| Material | Tech. | Dens. (g/cm³) | Young (MPa) | $M$ (MPa) | $\nu$ | $c$(m/s) | $c_{rel}$ | $z_{rel}$ | $\tau$ | $d$ ($\lambda$) | $d$ (voxel) |
|---|---|---|---|---|---|---|---|---|---|---|---|
| Somos® PerFORM | SLA | 1,61 | 10500 | 19830 | 0,32 | 3510 | 2,36 | 3,81 | 0,66 | 1,73 | 14 |
| Polyamida | FDM | 1,36 | 5000 | 10710 | 0,4 | 2810 | 1,89 | 2,57 | 0,81 | 2,12 | 17 |
| Standard Grey® | SLA | 1,16 | 3000 | 6970 | 0,41 | 2440 | 1,65 | 1,92 | 0,9 | 2,54 | 20 |
| PEEK | FDM | 1,32 | 3600 | 7710 | 0,4 | 2420 | 1,63 | 2,15 | 0,87 | 2,59 | 21 |
| PPS | FDM | 1,63 | 4000 | 8570 | 0,4 | 2290 | 1,54 | 2,52 | 0,81 | 2,84 | 23 |
| PEI | FDM | 1,27 | 3200 | 6380 | 0,39 | 2240 | 1,51 | 1,92 | 0,9 | 2,96 | 24 |
| VERO® Clear | SLA | 1,17 | 3000 | 4810 | 0,35 | 2030 | 1,37 | 1,6 | 0,95 | 3,73 | 30 |
| ABS | FDM | 1,09 | 2500 | 4420 | 0,37 | 2010 | 1,36 | 1,48 | 0,96 | 3,81 | 30 |
| PSU | FDM | 1,25 | 2610 | 4620 | 0,37 | 1920 | 1,29 | 1,62 | 0,94 | 4,4 | 35 |
| PLA | FDM | 1,26 | 2600 | 4170 | 0,35 | 1820 | 1,23 | 1,54 | 0,95 | 5,44 | 44 |
| Silicone | Mold | 0,98 | 150 | 1320 | 0,48 | 1160 | 0,78 | 0,77 | 0,98 | 3,56 | 28 |
| FEP | DLP | 2,17 | 500 | 1900 | 0,45 | 930 | 0,63 | 1,37 | 0,98 | 1,7 | 14 |
| TPU | FDM | 1,1 | 500 | 440 | 0,48 | 630 | 0,43 | 0,47 | 0,87 | 0,74 | 6 |

**Table 1**. Acoustical data of a series of 3D printable polymers, assuming for water $c=1485m/s$ and $z=1485krayl$. $\tau$ represents acoustic transmission on a single layer lens/water, nominal thickness, $d$, is obtained assuming a frequency of 760kHz and a voxel size of 0.244mm.

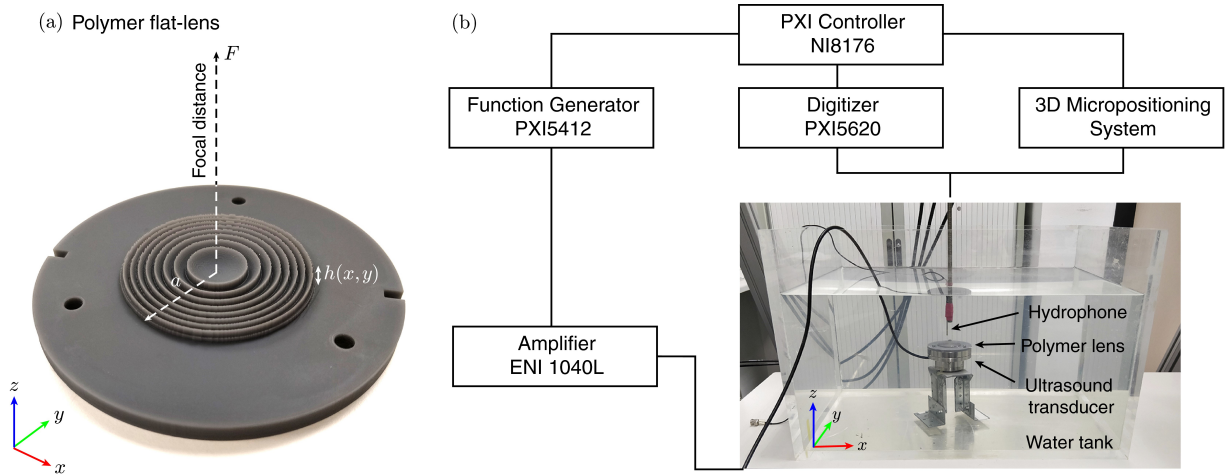

**Figure 2**. (a) Photograph of the flat-lens manufactured by stereolithographic 3D-printing technique using a photopolymer. (b) Scheme of the setup and equipment used for the experiments in water tank.

*2.6. Sources of errors in the printed lenses*

The main causes of error in the correction of the aberration, assuming a perfect application of the technique (i.e., disregarding any error associated with the position and orientation of the lens once printed), are described here. Even supposing an exact knowledge of the physical parameters at each point of the brain and skull and a set of equations accurately describing the dynamics of the wave, there would be two inevitable sources of error affecting the computational modelling of the lens: those relative to (i) the discretization of the domain and (ii) the theoretical simplifications assumed in the design of the lens shape starting from the registered hologram.

Regarding domain discretization, given the dependence between the accuracy of any numerical method and its computational cost, the only way to eventually achieve a null error would be with an infinite computational cost [46]. On the other hand, obtaining the shape of the lens from the registered hologram presents several difficulties. In the definition of the lens thickness as a function of the phase pattern recorded at the holographic surface -either by means of the Fabry-Pérot resonator approach [9] or by means of the proposed linear interpolation- the hypothesis of an idealized propagation of each ray in radial direction (transducer-center) from the transducer to the holographic surface is being assumed. This assumption is incorrect, since the ray changes direction at the periphery of the lens and not at the holographic surface; however, the thinner the lens, the truer it will be. The hypothesis assumed here of a linear relationship between the registered phase and lens thickness -rather than applying a more accurate model- is a second cause of error associated with this item.

In addition to the two aforementioned causes of error, exclusively associated with the numerical simulation that concludes obtaining the CAD model of the lens as an STL file, the process of generating a physical lens starting from the CAD model is affected by three new error sources: (i) inaccuracy of shape, (ii) discretization of the shape, and (iii) imprecision in the physical parameters that affect wave propagation, such as density, Young's modulus, Poisson's ratio, or acoustic absorption associated with the presence of anisotropy, porosity, heterogeneity, etc, related with the printing process. The inaccuracy of shape will be considered negligible in this study, since updated 3D printers can generate decimeter objects with inaccuracies of less than one tenth of a millimeter. As for the discretization of the shape of the lens, the dimensions of the elementary volume cell (voxel) depend on the 3D printing method (FDM, SLA, etc.) and the material used. Furthermore, given that 3D printing is a booming sector, the minimum voxel size is continuously being reduced. For the present study we have accepted a cubic voxel with a side of 0.244 mm, since it is a reasonable value that matches the grid element used in numerical calculations. Regarding the third cause of error, it should be noted that the anisotropy of the printing process itself (which is carried out in layers) can lead to large anisotropies in the value of the propagation speed, which is the most relevant parameter for the proper performance of these lenses. In this sense, it can be highlighted that if resins with fibers are used in FDM printing, we can find large anisotropies, such as compressibility modules of respective values $Bx = 10$ GPa and $B_{YZ} = 1,12$ GPa [41]. In general, SLA printing with subsequent curing using fiberless resins is the process that guarantees greater isotropy, although a certain degree of anisotropy or heterogeneity in the physical values of the printed lens can never be neglected [42-45].

Summarizing, five sources of error have been considered: discretization of the numerical domain, validity of the theoretical approach, inaccuracy in the form of the printed physical lens, discretization of the printed lens and imprecision in the relative p-wave speed of the printed lens. Between these, the inaccuracy in the form has been neglected and the size of the voxel associated with the printing process has been set to the size of the cell used for the numerical calculations. Thus three sources of error will be studied: spatial discretization, imprecision on the polymer p-wave speed and validity of the theoretical approach.

The effect associated with the imprecision in the relative speed of the printed lenses, accepted their isotropy, can be approached as follows. The maximum phase shift produced by a lens with a nominal thickness, $d$, being $c_{rel}$ the ratio between the p-wave speeds of the lens and the medium, is written

$$\beta = \left| \frac{2\pi}{\lambda_{water}} - \frac{2\pi}{c_{rel}\lambda_{water}} \right| d \tag{10}$$

If the lens is made with a negligible error in its dimensions, the absolute uncertainty of the phase shift, $\varepsilon(\beta)$, associated exclusively with the imprecision of the relative speed is obtained as follows

$$\varepsilon(\beta) = \frac{2\pi}{c_{rel}^2 \lambda_{water}} d \cdot \varepsilon(c_{rel}) \tag{11}$$

and knowing the value of the nominal thickness according to equation 9, the phase error as a function of the relative error of the relative velocity, $\varepsilon_r(c_{rel})$, can be expressed as

$$\varepsilon(\beta) = \left| \frac{2\pi}{c_{rel} - 1} \right| \varepsilon_r(c_{rel}) \tag{12}$$

Regarding the effect on the phase shift associated to the spatial discretization of the lens, we can state that the maximum error of the phase shift, $\varepsilon'(\beta)$, associated with the size of the voxel is written

$$\varepsilon'(\beta) = \left| \frac{2\pi}{n_d} \right| \tag{13}$$

Where $n_d$ represents the number of voxels associated with the nominal lens thickness. Thus, knowing the nominal thickness according to equation 9 and being $n_\lambda$ the number of voxels per wavelength in water, it is trivial to obtain

$$n_d = \left| \frac{c_{rel}}{c_{rel} - 1} \right| n_\lambda \ , \tag{14}$$

and combining the above equations 13, and 14 we get

$$\varepsilon'(\beta) = \left| \frac{c_{rel} - 1}{c_{rel}} \right| \frac{2\pi}{n_\lambda} \qquad , \tag{15}$$

For a better understanding of the implications of the expressions 12 and 15 the Figure 3 is attached. There can be appreciated that, if polymer p-wave speeds are close to that of water, the error associated with relative speed shoots towards infinity, whereas that associated with discretization tends to zero. It can be seen that the curves are asymmetrical, obtaining for both errors a more favourable trend in the high speeds zone.

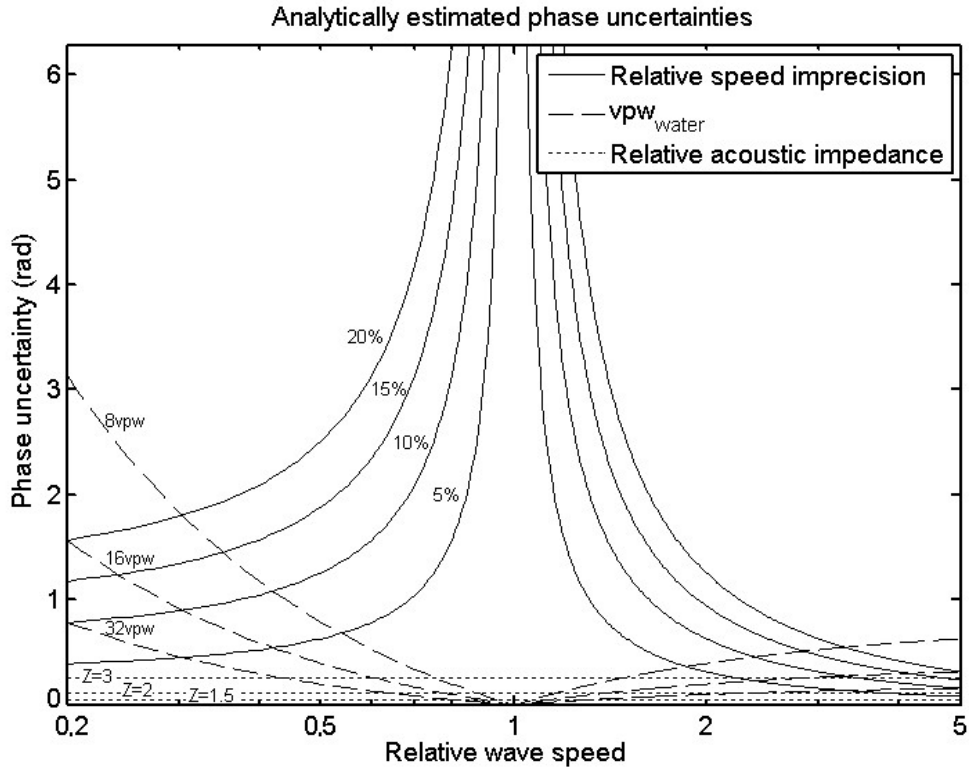

**Figure 3**. Analytical estimation of the uncertainties in the phase recreated by the 3D lens at the holographic surface associated with the discretization of the domain expressed in voxels per water wavelength, *vpw*<sub>*water*</sub>, (dashed), with the acoustic impedance of the lens when the definition of the thickness is approached by linear interpolation (dotted) and with possible differences between simulated and real values of lens p-wave speed (continuous).

The last error factor considered is the validity of the theoretical approach itself that is affected by an accidental and a systematic error. The accidental error is associated with the difference between the actual ray path and the idealized path that tends to coincide in very thin lenses. This accidental error is difficult to quantify, and will be evaluated through a series of numerical simulations. In addition, a systematic error, associated with the estimation of the thickness of the lens by means of a simple linear interpolation and not by means of a more accurate model such as the Fabry-Pérot resonator [9], is introduced. The reasons for using the linear hypothesis in this work are simple: First, (i) both models converge when the impedances of water and lens are equal and the total transparency is reached, (condition that has been applied in the set of numerical simulations evaluated), and second (ii) to apply the Fabry-Pérot resonator model, the acoustic impedances of the transducer, of the backing and of the coupling agent must be known. This would enforce us to play with many input variables with a minor effect on the quality of the approach, and therefore we have dismissed this variability for the study. The maximum value of the systematic error in the phase associated with the assumption of the linear approach is represented in Figure 3. This error has been calculated as the worst in the whole phase interval $[0, 2\pi)$ in the absence of coupling between transducer and lens by comparing the linear model with a Fabry-Pérot resonator.

Concluding, it can be stated that while the p-wave speed of each polymer affects lens shape and therefore the validity of the thin lens hypothesis, its acoustic impedance is not a direct cause of error in the phase generated at the registration surface. However, if this impedance is very different from that of water, transmission is decreased what is energetically inefficient. No study on optimal impedance has been proposed, therefore, as it is simple to conclude that the ideal lens would have an impedance equal to that of water.

## 3. Experiments and simulations

In order to determine the suitability of the polymers attending to their p-wave speed, a set of numerical simulations and a validation experiments are performed whose details are described here. The obtained results are presented and discussed in later sections.

### 3.1. Numerical Simulations

The simulations are carried out with or without skull interposed between transducer and target, which are placed at different points (P1, P2, P3) always between the spherical transducer and its geometrical center (Fig. 1). At each of these points and for each emitted frequency, we assume the same protocol: First of all, the phase pattern is recorded (by means of time reversal simulation from target to transducer) on the holographic recording surface, both with and without interposed skull. Then, in order to determine the characteristics of the ideal focus, the "gold standard emission", which consists on emitting from the registration surface affected by the phase pattern recorded without skull, is simulated. Next, a series of six lenses for six different speeds slower than water and another series of six lenses for six speeds faster than water are numerically generated from both the phase patterns recorded with or without skull. In the graphs and for the shake of brevity, we name concave or convex lenses the lenses produced respectively with fast or slow materials, since all the target points evaluated are located between the lens and its geometric center. The p-wave speeds evaluated ranges from one fourth to four times the wave speed of water.

Once calculated the lens at each configuration (i.e. at each point, at each frequency, with or without skull) and for each p-wave speed of the lens polymer, the numerical medium is modified by adding each Cartesian model of the acoustic lens placed in the correct position to generate the holographic pattern. Then, the time forward emission is simulated from the spherical transducer, emitting with uniform initial phase and amplitude, and placed concentric to the phase registration surface at a distance from it equivalent to the value of the nominal lens thickness. A backing with the same acoustic properties of the lens is added to avoid the numerical issues related with placing the emitting nodes at a boundary.

The density of each simulated lens is established in accordance to its p-wave speed in order to set a unit normalised impedance for each polymer. Therefore, the simultaneous evaluation of velocity and impedance effects is avoided, but the properties of the simulated polymers could not correspond to any physical material.

For each of these simulations, the quality of the -3 dB and -6 dB focus beams is evaluated attending to seven quantitative indicators described below. The particular values used in the simulation series are the following: the p-wave speeds of the fast lenses are respectively 4, 2.5, 1.75, 1.5, 1.375 and 1.3 times the speed of water, whereas the p-wave speeds of the slow lenses are obtained dividing the speed of water by the same series of numbers; the spherical transducers have a radius of 59 mm and an aperture of 64mm for configuration P3, and 30mm for P1 and P2. Figure 1 shows the target points whose respective distances to the holographic surface are 24.2 mm 13.8 mm and 30 mm. In short, there are 96 time forward simulations for the small aperture lens (two types of lens, two frequencies, with and without skull, two positions, and six speeds) and 13 simulations for the lens of larger aperture.

### 3.2. Quantitative focusing indicators

To assess the quality of the focal spot achieved by each kind of material we calculate seven quantitative indicators as described by Ferri et al. [8] for both -3 dB and -6 dB focal beams. These indicators evaluate for each focal beam (i) its longitudinal and transverse deviations from the target point ($z$, $R$), (ii) its transverse an total gyradii ($k_R$, $k$), (iii) orientation, $\Delta\varphi$, (iv) focal volume, $V$, and (v) energetic overlapping with the ideal focus, $I_i\,(\%)$.

### 3.3. Experiments

A validation of the simulation was performed by measuring the acoustical field of a 3D printed test lens in the ultrasonic regime. The experiments were done inside a 1 x 0.75 x 0.5 m degassed-distilled water tank at 21º C, as shown in Figure 2(b). The lens was excited using a custom-made ultrasonic source composed of single-element piezoelectric transducer mounted in a custom designed stainless-steel housing with a diameter $2a$ = 50 mm. The transducer was driven with a 50 cycles sinusoidal pulse burst at a frequency of $f$ = 1.112 MHz. The pressure field was mapped by a 500-μm needle hydrophone. The hydrophone signals were digitized at a sampling rate of 64 MHz averaged 50 times to increase the signal to noise ratio. A precision 3D micro-positioning system was used to move the hydrophone in three orthogonal directions with an accuracy of 10 μm. Scanned area covered from -3 mm to 3 mm in the $x$ and $y$ axis, and from 10 to 35 mm in the $z$ axis, using a step of 0.1 mm in all directions. Temperature measurements were performed throughout the whole process to ensure no temperature changes of 1º C.

## 4. Results

This section presents the results of the set of simulations evaluated and the validation experiment carried out. Numerical parameterization of the focusing at each configuration simulated has been performed for both -3dB and -6dB beams, however, given the similarity of the results obtained; only the indicators obtained at -6dB will be presented.

### 4.1. Emission in water at 760 kHz

**Positional deviation of the focus:** The longitudinal focal point deviations of the -6 dB beam, for the simulations performed with each of the lens evaluated, are shown in Figure 4(a). It can be seen that the trends found in the deviation values are different for concave and convex lenses, but somehow independent on the evaluated point. The obtained longitudinal deviations for the convex lenses seem to present optimum values for nominal lens thickness between 10 and 20 voxels whereas, for concave lenses at points P1 and P2, the results show that the thinner, the better. Concave lens at point P3 exhibits an intermediate behaviour, with a flat trend between 10 and 15 voxels.

The transverse deviations found (Figure 4(b)) are negligible compared to the emitted wavelength ($\lambda_{water} \approx 2$ mm at 760 kHz) and to the size of the voxel for all the configurations evaluated, showing a maximum value of 60 μm which is less than one third of the computational internode distance. No trend can be analyzed as the calculated transverse deviations are simply showing an stochastic behaviour.

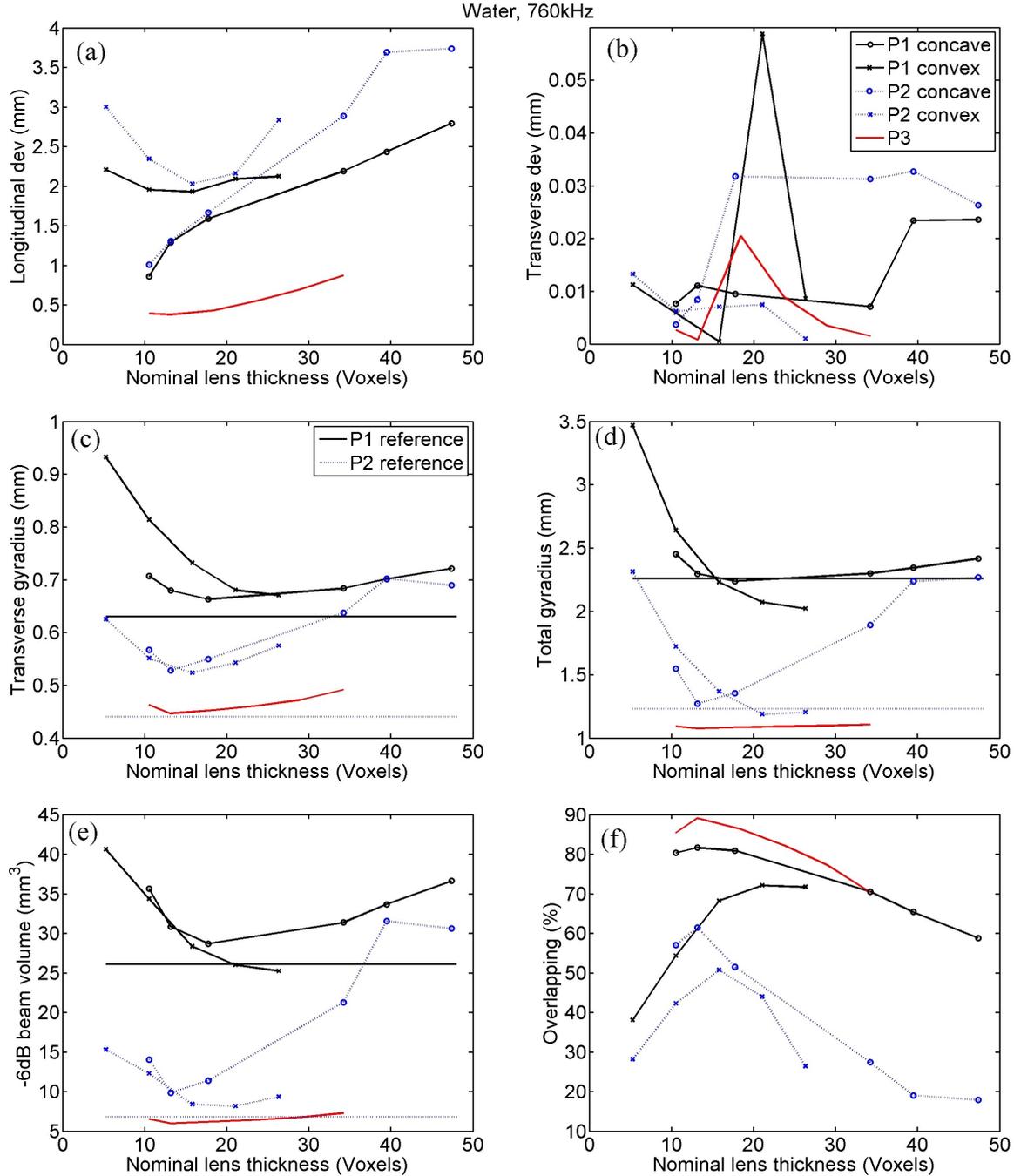

**Figure 4.** Focusing quality indicators calculated by simulation in water with 760 kHz, at points P1, P2 and P3, for concave (fast) and convex (slow) lenses. Indicators are: (a) Longitudinal and (b) transverse deviation, (c) transverse and (d) total gyradius, (e) beam volume and (f) energetic overlapping.

**Gyradius:** The proper beam shape, considered similar to the shape of the reference beam, is evaluated using the transverse, $k_R$, and total, $k$, gyradii of the -6 dB focal beams (Figures 4(c-d)). The reference -6 dB beam obtained at P1 has the values $k_R = 0.63$ mm $k = 2.26$ mm, and at P2 the values are $k_R = 0.44$ mm $k = 1.23$ mm. Both transverse and total gyradii at the -6dB focal beams have certain similarities with those of the reference beam, at all the evaluated points and for concave and convex lenses. Particularly, and with the exception of a couple of cases of the total gyradius for convex lenses at both P1 and P2, the beams with interposed lenses are larger than the reference beam, as could be expected. Therefore, the smaller the gyradius the more accurate the beam. Then, we found that concave lenses –as in positional deviation- show its best results for nominal lens thickness between 10 and 20 voxels at both P1, P2 and P2; whereas convex lenses seems to present the best values with lightly thicker lenses for P1. At P2, however, the transverse gyradius presents a clear trend with an optimum between 10 and 20 voxels and the trend of the total gyradius behaves pretty like in P1. Concave lens at P3 shows relatively flat trends but an optimum range between 10 and 20 voxels can again be appreciated.

**Focal volume:** The focal volume is compared with that of the reference beam founding that reference is always smaller with values 26.1mm3 for P1 and 6.8mm3 for P2. It is somehow obvious that the focal volume is highly linked with the gyradii values, so the trends found in Figure 4(c-e) present high similarities. As in gyradius, we find that for concave lenses at the three points the trends signals clearly an optimum when the nominal lens thickness is between 10 and 20 voxels, and in convex lenses the optimum for both P1 and P2 is found at slightly larger lens thickness

It is also remarkable that the most accurate values present a great similitude with the reference beam, with differences as slight as 1% whereas the thickest or thinnest lenses presents values with differences with the reference beam as large as 56%.

**Orientation** No plot of orientation is presented, as their average values in water are smaller than 0.009rad (0.5º). This indicator here only informs about numerical uncertainty as in the case of transverse deviation.

**Energetic overlapping:** Figure 4(f) shows that the largest overlappings for -6db beams are achieved at point P3 and P1 with concave lenses, with values as big as 89%. Concave lenses exhibit a slight better behaviour than convex ones for both points P1 and P2, and P1 shows better overlapping than P2 regardless the kind of lens. The trend at the five curves shows similarities with that observed in previous parameters such as focal volume and gyradius. The best overlappings are found with lens thickness between 10 and 20 voxels for all the cases with the only exception of convex lenses at P1where the optimum is found between 15 and 25 voxels.

*4.2. Transcranial emission at 760 kHz*

**Positional deviation of the focus:** In transcranial emission both longitudinal and transverse focal point deviations (z and R) are relevant, and its values for the -6 dB beams are shown in Figure 5(a,b). Transverse deviations obtained are relatively small compared to the emitted wavelength ($\lambda_{water} \approx 2$mm at 760kHz) and to the size of the voxel in the evaluated cases, for each of the thicknesses, points and types of lens. Their values smaller than 0,6mm in all the cases are notably smaller than the values of the longitudinal deviation, that are smaller than 3 mm in all the cases.

Despite the small values found for transverse deviations the trends can be easily appreciated, and are notably similar those found in previous indicators such as gyradius or focal volume in underwater case. Again the smallest deviations for concave and convex lenses at the points P1 and P2 are obtained in the interval between 10 and 20 voxels; point P3 exhibits an aggressive trend with a minimum around 15 voxels. The smallest deviations at P1 are achieved by the concave lens and, at P2, by the convex one.

For longitudinal deviation, there are some similitude between the trends obtained for the underwater and the transcranial case. In transcranial case, trends are less explicit, but show again that the optimum nominal thickness for the convex lens is found between 10 and 20 voxels at P1 and bigger than 10 voxels at P2, whereas, for concave lenses, at points P1 and P2, it is found that the thinner the better, and at point P3 the optimal interval is again between 10 and 20 voxels.

**Gyradius:** Transverse, $k_R$, and total, $k$, radii of gyration of the -6dB focal beams are shown at Figure 5(c,d). A great concordance between the data at both plots can be highlighted for points P1 and P2, since the trends of transverse gyradii are almost exactly proportional to those of the total gyradii for any type of lens. On the contrary, point P3 shows transverse gyradii values notably larger than expected. The values at the three points, for any type of gyradius at any simulation, are again slightly bigger than the reference values, displayed at Figure 4(c,d). It is notable the similitude between the results found with concave and convex lenses, for both P1 and P2. However, it is difficult to find conclusive results about the preferable nominal thickness, as the curves are quite flat; then, at point P1 we could appreciate that the smallest gyradii with convex lenses are in the interval between 15 and 25 voxels, but in the rest of the cases we just can affirm that this interval is not worst than any other

Finally, if we compare these data with those found for the underwater case, we can appreciate that the transcranial aberrated corrected beam is slightly bigger indicating that the dispersion of the energy in the skull is not absolutely compensated by the lens. This effect is particularly appreciable in the transverse gyradius at point P3.

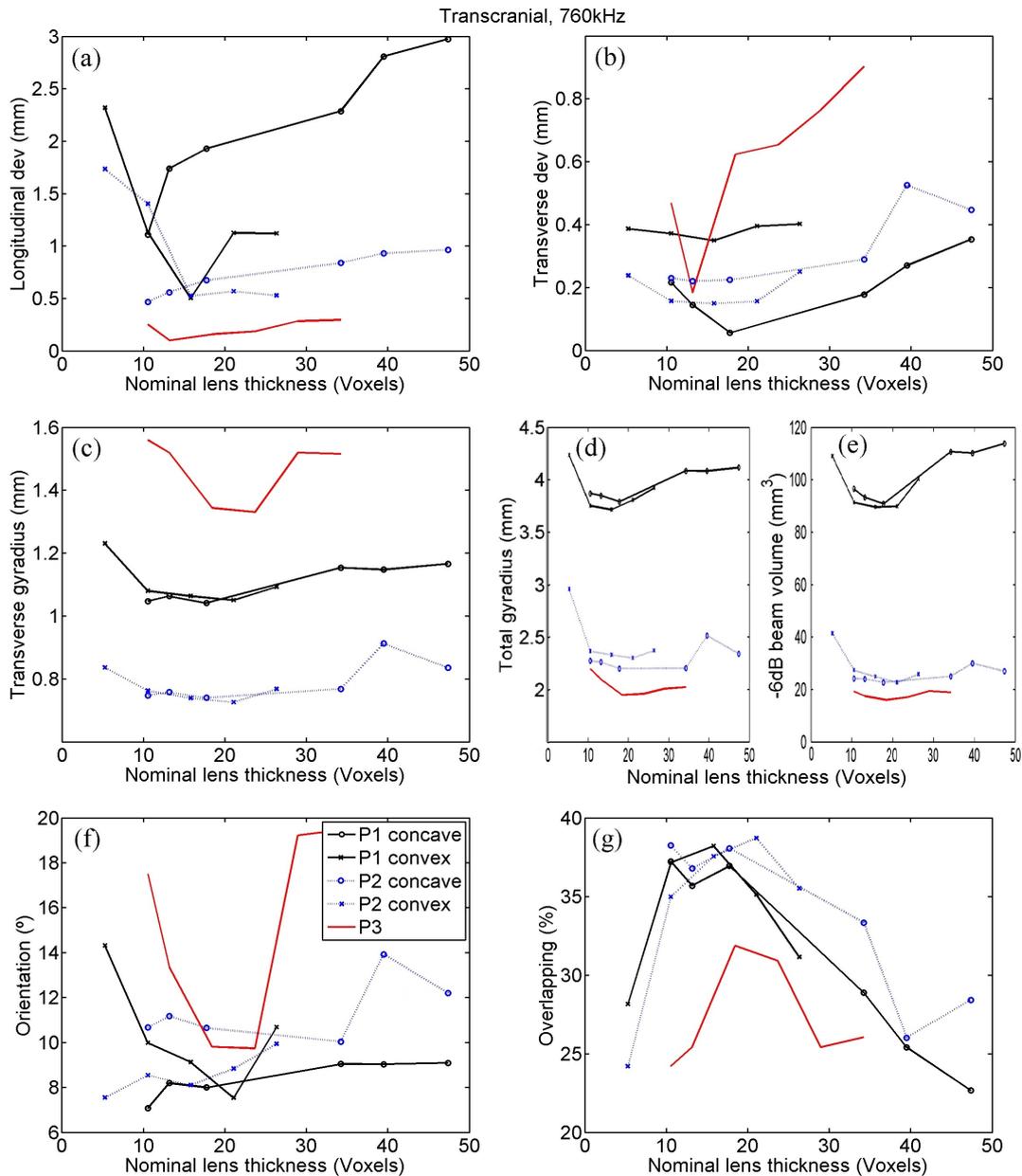

**Figure 5.** Focusing quality indicators calculated by transcranial simulation with 760 kHz at points P1, P2 and P3, for concave (fast) and convex (slow) lenses. Indicators are: (a) Longitudinal and (b) transverse deviation, (c) transverse and (d) total gyradius, (e) beam volume, (f) orientation and (g) energetic overlapping.

**Focal volume:** In brief, we can summarize that the results found for the focal volume are in great accordance with that of gyradius (Figure 5e). So, convex and concave lenses present similar results between them, and the trends found in underwater and transcranial cases are also very similar. What can be highlighted is that the aberrated corrected beams are notably bigger that those found in the underwater case. This difference could seem excessive if compared with that found at radii but this apparent incoherence is associated with (i) the different units of gyradius and volume so the relative uncertainties in volume are three times bigger than those of gyradius, and additionally (ii) the central part of the beam is reasonably similar between transcranial and underwater cases, but as we separate from the target point there is more diffuse energy in the transcranial case, and an important fraction of this diffuse energy exceeds the arbitrary threshold of -6dB. Finally the interval of lens thickness between 10 and 25 voxels is associated with the best results at P1 and this fact is not in contradiction with the flat trends found at P2 and P3.

**Orientation** This indicator, as has been defined, is highly sensitive to small asymmetries or irregularities associated to a non exact aberration correction. Therefore we can see trends much more abrupt than in other evaluated indicators (Figure 5(f)). With the exception of point P3, with optimum values at the range between 10 and 25 voxels, there is not a clear interval of preferred nominal thicknesses, but at least we find again that the worst values are at the extremes of the curves, what is somehow coherent with the rest of the evaluation. No clear preferences are obtained between concave and convex lenses, or between further and closer points.

**Energetic overlapping:** The values of this indicator are deeply linked to the values of previously commented ones, and therefore data at Figure 5(g) summarises what we have seen. Then, in comparison with overlappings at the underwater case, we find here smaller values and rougher trends. Particularly at point P3 we obtain the worst focusing in transcranial whereas was the best focused in water, being this great reduction of focusing probably related with the irregularity of the skull region crossed in this case. All the overlappings are between 20% and 45%, and the interval of nominal thickness in which these values are largest is again between 10 and 25 voxels for both concave and convex lenses at the three points evaluated. Finally, the values obtained for this indicator in transcranial simulations seem to be notably independent on the type of lens or the target distance.

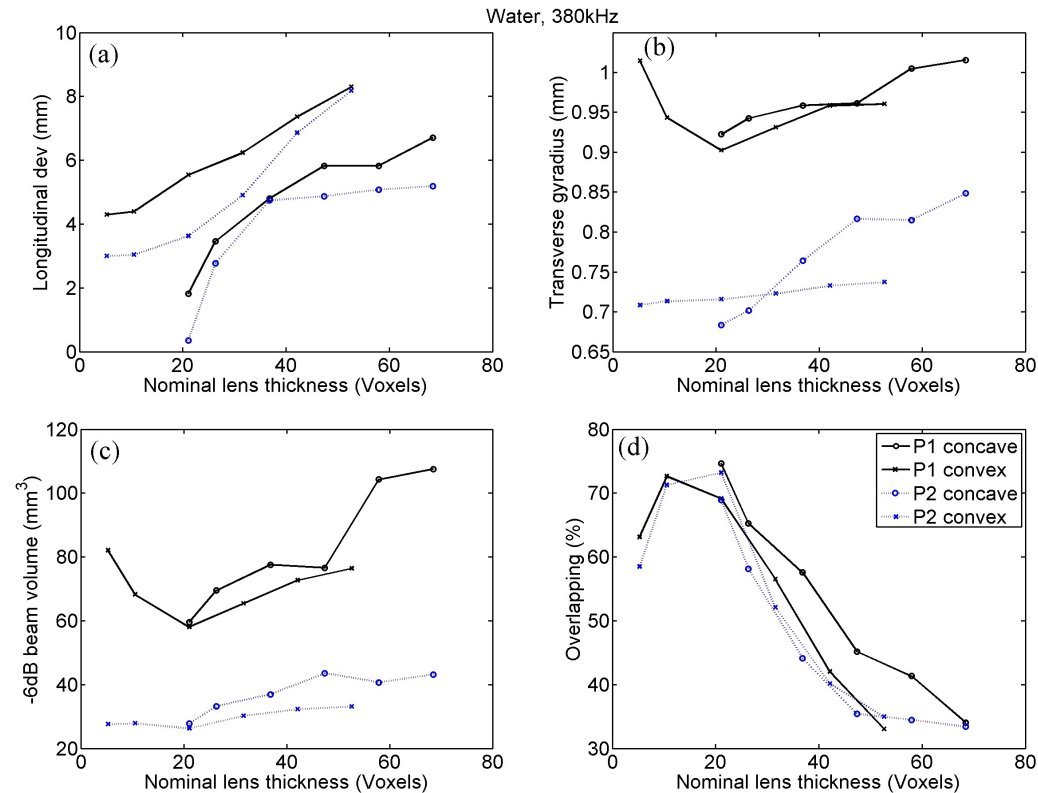

**Figure 6.** Focusing quality indicators calculated by simulation in water with 380 kHz, at points P1, and P2, for concave (fast) and convex (slow) lenses. Indicators are: (a) Longitudinal deviation, (b) transverse gyradius, (c) beam volume and (d) energetic overlapping.

### 4.3. Emission in water at 380 kHz

Since the results found are very similar to those seen at 760kHz, the analysis of the graphical information shown in Figure 6 is here faced as a whole. Thus, we can point out that, in the case of concave lenses, all the parameters in both points show the same type of trend, consisting of a worsening of the indicator quality as the thickness of the lens increases. This is not contradictory with what was seen in the evaluation at 760kHz, but is due to the fact that the smallest possible lens thickness -conditioned to the value of the water wavelength in the concave lenses- is greater than the values that define the optimum interval. In the case of convex lenses, whose thickness is not limited a priori, two types of trend are shown in the indicators and one exception. One type of trend, appreciated at beam volume and at both gyradii, is simply flat, so that no conclusion can be drawn from it. For the rest of the indicators, and for both points, the same trend seen for the previous frequency is observed, that is, we find an area of optimum values between 10 and 30 voxels thick. Although this trend is slight in several indicators, it is clearer for overlapping, which is the most relevant indicator. Finally, it is worth mentioning that (i) in the transverse deviation no significant trend is observed, since its values in the underwater case are only indicative of the numerical error and that (ii) in the longitudinal deviation no clear optimum is appreciated, but we interpret that the indicator improves the smaller the thickness of the lens. This last result could be justified by the fact that the thickness of the lens conditions the position of the transducer to maintain the position of the phase registration curve. Therefore, it seems logical that the change of position of the transducer in the longitudinal axis affects the longitudinal position of the focus, so that this indicator is the only one in which the results indicate that it is preferable to reduce the thickness as much as possible.

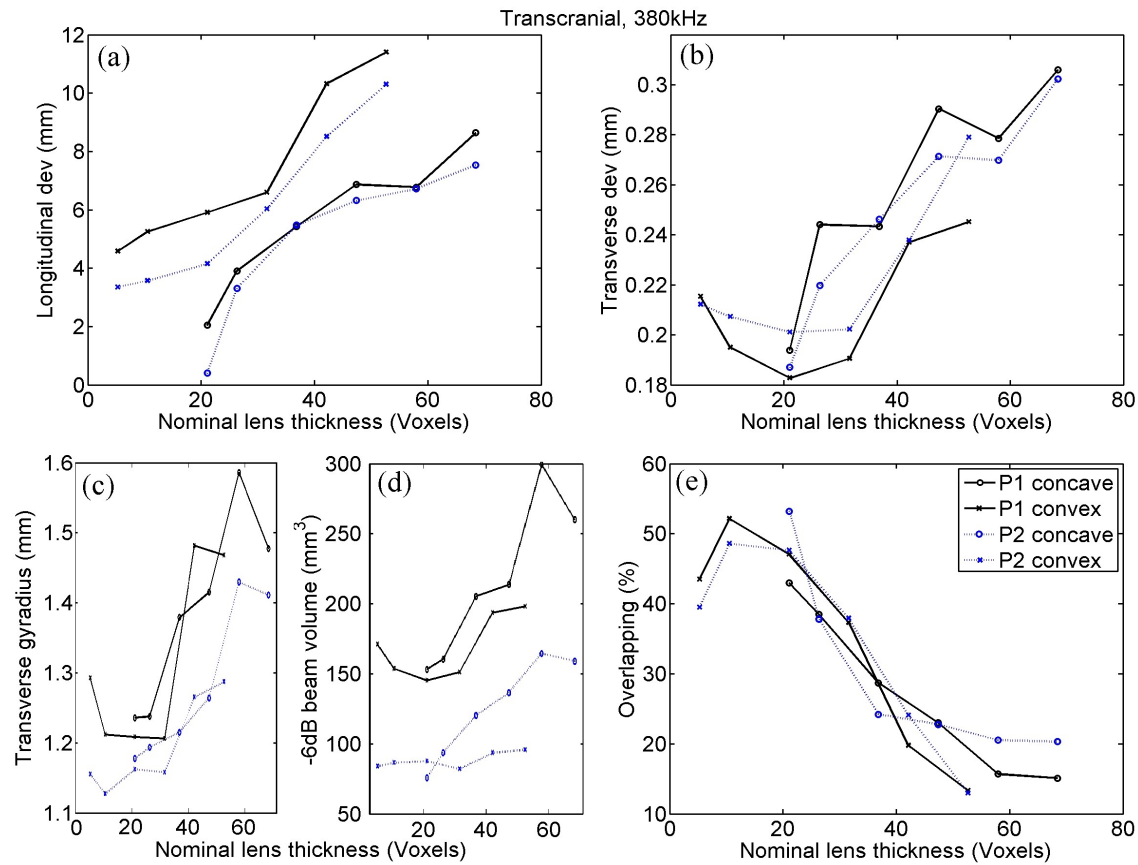

**Figure 7.** Focusing quality indicators calculated by transcranial simulation with 380 kHz at points P1 and P2, for concave (fast) and convex (slow) lenses. Indicators are: (a) Longitudinal and (b) transverse deviation, (c) transverse gyradius, (d) beam volume and (e) energetic overlapping.

### 4.4. Transcranial emission at 380 kHz

The results found (Figure 7) are very similar to those of the underwater case at the same frequency, although the transverse deviation, which is not a significant indicator in the underwater case, broadly shows the same trend seen in the rest of the parameters. That is, an area of optimum values between 10 and 30 voxels for convex lenses, while in concave lenses, which do not cover this range, the indicator worsens as the thickness increases. It is also worth noting the high parallelism of the results found in both underwater and transcranial cases, which can be justified because the capacity of the skull to generate aberrations is lower in the case of a larger wavelength

### 4.5. Experimental validation

The results of the field measurements are summarized in Figure 8. First, the measured field and the corresponding simulation at 1.12 MHz along the axis of symmetry of the lens are shown in Fig. 8(a). A good agreement is found between the simulation and the experiment. The focal spot is well described by simulations. Note that the focal spot peaks at z = 28.3 mm instead at z = F = 30 mm. This is due to the diffraction effects of the wavefront, in accordance with the axial-focal shift expected in any focused source as described in [47]. The pressure distribution was measured in the sagittal plane, $p(x, z)$, and the normalized absolute value of the field is shown in Fig.8(b) for the experiment and Fig.8(c) for the simulation. An excellent agreement is found between both fields. Remark that the field shows excellent symmetry as expected by the geometry of the lens. The corresponding transverse field distribution measured at the focal spot is shown in Figs. 8(d-f) for both, simulation and experiment. The focal spot presents a sharp transverse size smaller than the wavelength (about 1.2 mm). Although small discrepancies are observed at the side-lobes of the pattern, the sharp focus observed experimentally agrees well with simulations, showing that the proposed computational method describe in an accurate manner the expected behavior of the holographic lenses.

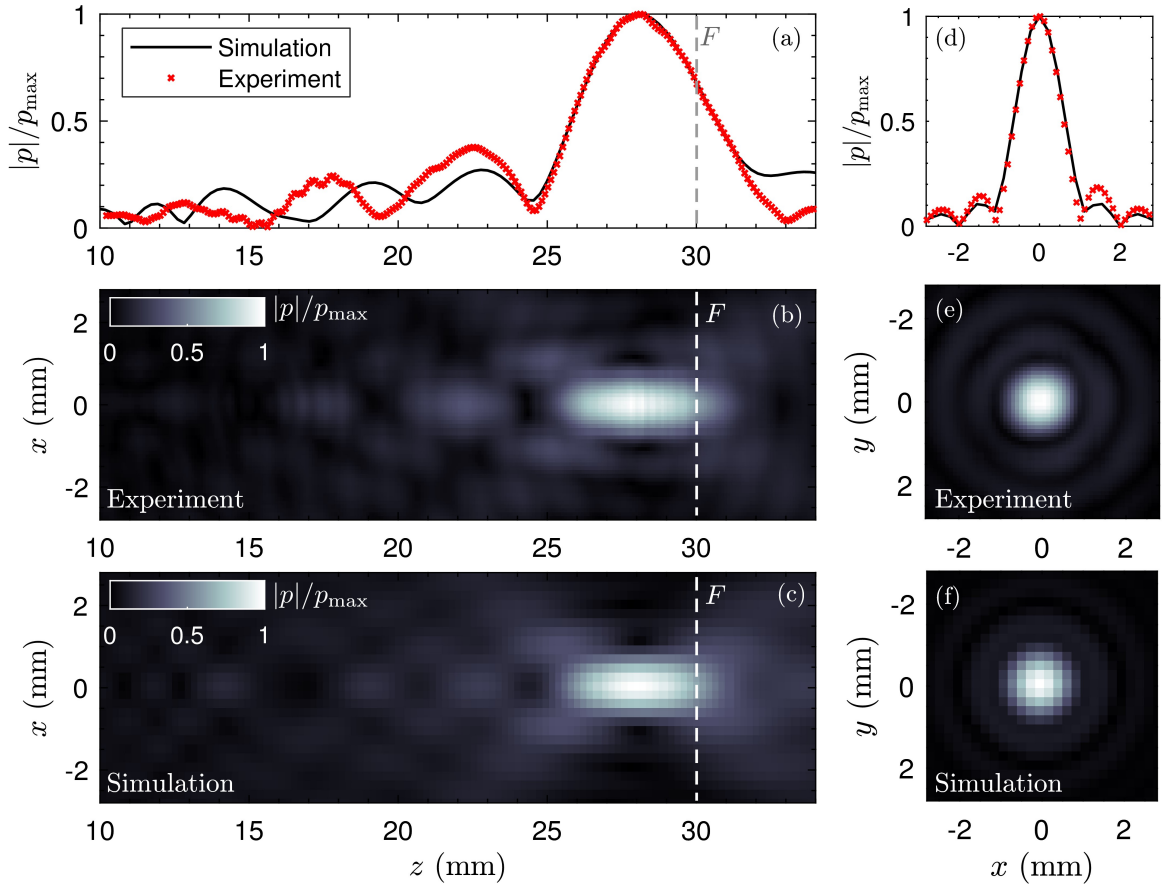

**Figure 8**. Results of the experimental validation test. (a) Experimental (red markers) and simulated (black continuous) normalized pressure field distribution measured at the axis of symmetry. Pressure field distribution in the sagittal plane, $p(x, z)$, obtained (b) experimentally and (c) numerically. (d) Experimental (red markers) and simulated (black continuous) normalized pressure field distribution in the transverse direction $x$ measured at the focal spot. Corresponding pressure distributions in the transverse plane, $p(x, y)$, obtained (e) experimentally and (f) numerically.

### 5. Discussion

Although the extension of this study does not allow the quantification of the statistical significance of the results obtained, certain relevant evidences can be extracted, which are discussed below. In order to obtain a conclusive result regarding the precise physical parameters that define the optimum lens, it would be necessary -in addition to more similar experiments to increase the statistical significance of the results- also to increase the number of types of simulations, with more wavelength values, more distances to the focus point (including points beyond the geometric centre of the transducer) or more lens apertures. Various numerical methods should also be applied, as the wave speed directly affects the stability of the calculation which limits the reliability of the results for lenses with speeds highly different to the medium, whether faster or slower.

However, although we cannot determine the exact properties of the optimal lens, the presence of a trend that allows the establishment of an interval of desirable values has been clearly evidenced by the results obtained. Indeed, as seen in Figures 4 to 7, all the quantitative focusing indicators obtained in the simulations with 760kHz, for the two types of lens, at any point and with or without skull, present a behaviour regarding the lens thickness that can be expressed as follows. For thicknesses greater than a certain value of about 15 or 20 voxels the indicators worsen with the increase in thickness, although this relationship is moderate and sometimes even flat (as in the case of the transcranial focal volume in point 2). For thicknesses smaller than 10 or 15 voxels the dependence between the value of the indicators and the thickness is much more abrupt, so that thicknesses under 10 voxels are clearly discouraged. This trend is manifested in all indicators in the case of slow lenses, since with fast lenses thicknesses smaller than water wavelength cannot be obtained, however, this result could not be conclusive because the bad value of indicators could be partially associated with the numerical error in the propagation inside the lens, given that the number of points per wavelength in extremely thin lenses is less than recommended [32]. In any case, and due to the fact that the impedance value has been fixed to achieve total transparency, which means that there are no internal reflections in the lens, we can assume that the possible numerical error in the phase associated with a path inside the lens as short as 5 or 10 voxels would not have a considerable cumulative effect, but that the origin of the error is precisely the strong discretization of the values of the phase at the exit of the lens associated with the fact that its thickness is defined by such scarce number of voxels. This hypothesis is reinforced by the fact that in the range of 10 to 15 voxels we can already see, albeit in a moderate way, that lower thickness implies worse value of the indicators.

Similar conclusions can be drawn from the analysis of the results obtained with the 380kHz emission. Fast lenses here are always thicker than the optimum, so it is observed that greater thicknesses directly imply worse indicators; as for slow lenses, we observe the trend commented in the case of 760Hz with all indicators, with the sole exception of the longitudinal deviation.

In the simulations at both frequencies certain additional relationships can be seen, which are briefly discussed below. Thus, as point P1 is farther away than P2, higher values of focus size or gyradius are obtained, which is predictable; however obtained values of the transverse deviation or the transcranial energetic overlapping are similar at both evaluated points. In water simulations, however, it seems clear that the indicators are better for the farthest point, which makes sense since a farther target implies smoother lens curvatures which results in (i) the appearance of fewer edges associated with the phase shift 2π; and, on the one hand and, (ii) a smaller deviation of rays at the boundary of the lens, which is better suited to the thin lens model and additionally activates less energy in form of s-waves. Another foreseeable result is that in the process of passing through the skull, even if the incidence is made with phase correction, energy is dispersed, so that the beam contours are diffuse and indicators obtained are generally worse than that of underwater case.

The simulation with several frequencies -one of which deviates from the usual values in medical treatments- aims to study whether the optimal thickness values are intrinsic or dependent on the wavelength, given a voxel size. It is observed that with different wavelengths the curves are not displaced in the axis of thicknesses, but that, for both 380kHz and 760kHz, the best indicators are in the zone between 10 and 25 voxels thick, regardless of the kind of lens, the distance transducer-target, or the presence or absence of skull. This result deserves to be highlighted by its relevance in order to generalize the study, as all the magnitudes measured in meters of the problem are related. Therefore, if, for instance, the lens aperture, its radius, the transducer-target distance and the wavelength were duplicated simultaneously, then we would have a mathematically equivalent situation to the initial state (disregarding the effect of absorption) but with a voxel of half the size. If we change only some of those lengths, we are facing different cases. With that said, finding that the ideal range of thickness (in voxel) does not depend on either frequency or wavelength is a remarkable result. In this same sense, it is worth mentioning that the frequencies used in medical treatments (about 1MHz) are associated with an optimal lens thickness that can be achieved with polymers of fairly common properties, both in the case of concave and convex lenses, as can be seen in the Tables 1 and 2. On the other hand, in the case of 380kHz, not usual in transcranial FUS, it can be seen (Table 2) that no material has the necessary rigidity to reach the concave lenses optimum thickness since fast lenses have a nominal thickness larger than the wavelength of the sound in the water. This could be solved with slow lenses whose thickness is not limited, although, as will be seen below, slow lenses are affected by other inconveniences.

The error associated with the possible difference between the speed of the printed lens and the one assumed in the simulation -represented in the analytical curves of the Figure 3- has not been evaluated by the set of simulations. This error presents a notable asymmetry, so that it tends asymptotically to zero on the side of fast lenses, whereas on the side of slow lenses (elastomers) both this error and that related to discretization tend to not negligible values. Thus, from the point of view of the focusing quality, we could conclude that lenses "as slow as possible" are discouraged a priori while lenses "as fast as possible" are recommended as long as they have an adequate impedance, as will be discussed below.

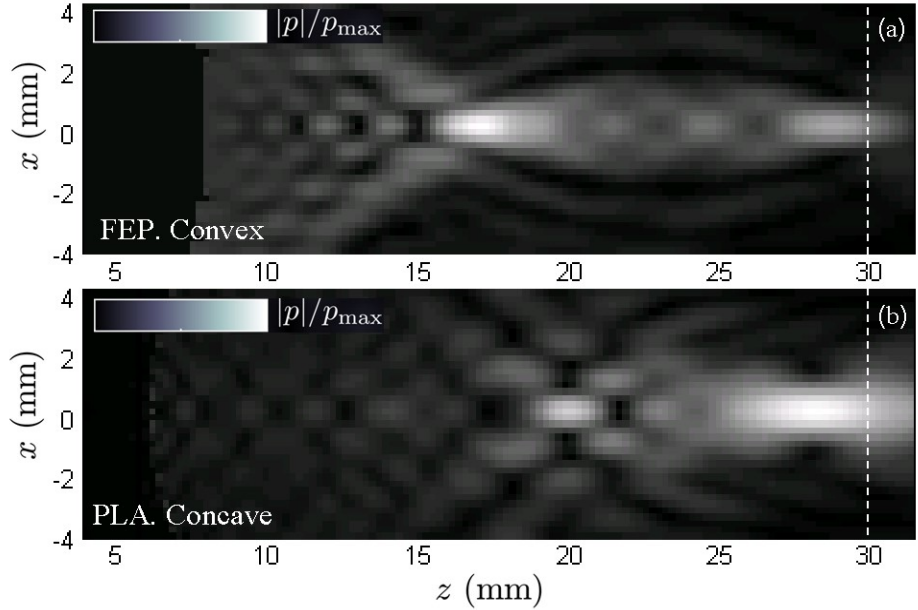

**Figure 9**. Pressure field distribution in the sagittal plane, p(x, z), obtained numerically from two lenses generated from a particular registered pattern: (a) convex lens made of fluorinated ethylene propylene and (b) concave lens made of polylactic acid.

New evidence in favour of rigid lenses can be drawn from the preliminary results of the experimental validation (Figure 9); its analytical basis is now discussed. Most of the energy is propagated in the form of p-waves, however, s-waves cannot be neglected due to the abrupt curvature often presented by the lenses. Thus, the fact that the s-wave speed is always lower than the p-wave speed affects each type of lens very differently. In slow lenses, the s-wave speed is even slower which makes the s-waves even more refractive, allowing the generation of secondary foci between the primary and the lens; these foci, by geometric proximity to the source, could eventually concentrate high energy which is undesirable. In fast lenses, on the other hand, the value of the s-wave speed can be faster than the sound speed at water but also slower. In the first case, the refraction would be lower than in p-waves so that -if a secondary focus were produced- it would be beyond the primary focus accumulating low energy; in the second case (s-wave speed lower than water) the s-wave would present an opposite vergence to the p-wave, so the lens would disperse the energy instead of concentrating it. Both possibilities in fast lenses are clearly favourable with respect to the risks associated with the application of slow lenses. However, if these are made with elastomers with a Poisson's ratio very close to 0.5, the medium behaviour is practically equivalent to a fluid with a single associated propagation speed, which would somehow validate the use of slow lenses.

The set of simulations aimed to find the range of preferred thicknesses accounting exclusively the effect of the polymer wave speeds. However, the important effect of the impedance must be highlighted and it is discussed briefly here. From the theoretical point of view, it is simple to establish that its preferred value is equal to that of water, but in practice this cannot always be achieved. In fact the velocity of the p-waves, $c_p$, and the acoustic impedance, $z$, are related according to the expression $z = \rho_o c_p$, where $\rho_o$ is density. Thus, as shown at Table 1, whereas in fast materials the density is practically constant so that an increase of the speed carries implicitly an increase in the impedance; in the slow ones one can find different densities and therefore water normalised acoustic impedances close to the unit. Highest transparencies nearing 100% are found with fluorocarbons, such the fluorinated ethylene propylene (FEP). Therefore, from the point of view of energy efficiency, it would be preferable a priori to use slow lenses as long as they are elastomers with unitary water normalised impedances. However, from a strictly focusing-quality point of view, the use of fast lenses is still recommended because, as stated in previous sections about sources of error, by knowing the physical properties of the transducer and backing, the phase error associated with internal reflections in the lens can be corrected completely. In any case, although the phase error associated with impedance is known and can be corrected -by designing the lens using Fabry Perot's hypothesis [9]- the fact that total transparency is only reached when impedances of lenses and water are equal constitutes a clear handicap for rigid lenses.

An additional aspect to keep in mind is that the transmission from the transducer to the supposedly spherical inner surface of the lens will be notably affected by the fact that 3D printing does not produce a perfect curve but a voxeled Cartesian model. Thus, the contact between the transducer and the lens is not homogeneous but concentrations of stresses may occur in certain specific vertexes of the lens with the risk of high transmission irregularities. These concentrations will be more pronounced in rigid and crystalline plastics than in deformable and amorphous plastics such as elastomers. Although high impedance gels [48] can soften this effect, we recommend facing its quantitative study in later works. These arguments are no longer valid when emitting by flat transducers, as in the validation experiment.

The choice of a nominal lens thickness associated with a phase shift $2\pi$, not discussed in previous sections, can be justified attending to the results of the simulations. Indeed, they are coincident with the thin lens hypothesis assumed, since -disregarding the error associated with the spatial discretization of the lens- the finer the lens the better the focusing. Therefore, even when Fresnel lenses with a $2\pi$ phase shift have many areas of abrupt curvature; their use is reasonably justified compared to continuous lenses or lenses with a larger associated phase shift that would be much thicker.

| | | Experiment | Simulation | | | | Assumed data | |
|---|---|---|---|---|---|---|---|---|
| | Frequency /kHz | 1112 | 760 | 380 | | | $c_{water}=1485$m/s | |
| | $n_d$ | **15** | **15** | **15** | 28 | | voxel size $\quad dh= 0.244$mm | |
| | $d$ / mm | 3,66 | 3,66 | 3,66 | 6,84 | | | |
| | $d / \lambda$ | 2,74 | 1,87 | 0,94 | 1,75 | | Applied equation | |
| Fast lens | $c_{relative}$ | 1,57 | 2,14 | --- | 2,33 | | $c_{LENS} = \dfrac{c_{H_2 0}}{1 \pm \dfrac{c_{H_2 0}}{f \cdot n_d \cdot dh}}$ | |
| | $c$/(m/s) | 2340 | 3180 | --- | **3470** | | | |
| Slow lens | $c_{relative}$ | 0,73 | 0,65 | 0,48 | 0,64 | | | |
| | $c$/(m/s) | 1090 | 970 | 720 | 940 | | | |

**Table 2.** Recommended p-wave speeds and nominal lens thicknesses, $d$, for the experiment and simulation frequencies, assuming that the number of voxels associated with the ideal nominal thickness value, $n_d$, is 15. Last column show the minimal value of $n_d$ that can be achieved by the fastest polymer listed in Table 1.

Summarizing, we could say that slow lenses have the advantages of (i) certain independence between p-wave speed and acoustic impedance so that it is possible to achieve simultaneously speeds very different to water with very similar impedances and therefore high transparency; and, (ii) in the case of lenses made of elastomers to be used with spherical transducers, an easier mechanical coupling that avoids concentrations of tensions in the micro-vertices of the Cartesian mesh. Fast lenses, on the other hand, have the advantage of (i) a lower risk of energy concentrations in secondary foci associated with s-waves, as well as (ii) a much less vulnerable behaviour related to possible inaccuracies in the manufacturing processes of the lens. Finally, it should be noted that results of simulations indicate that the range of optimum lens thicknesses given a 3D printer voxel size is coincident for both fast and slow lenses, and notably independent of other factors such as emission frequency, transducer aperture, or transducer-target distance.

## 6. Conclusions

In this work we have looked for the best polymers to 3D print holographic acoustic lenses, in particular to be applied in the correction of transcranial FUS aberrations. Therefore, although there are many parameters affecting the focusing quality of a lens, some of them affect in a trivial way, being simple to establish a priori the desirable value. We can, for example, establish as desirable that the material must be homogeneous and with low acoustic absorption. However, there are two main parameters with a non-trivial relationship with quality: the impedance and the p-wave speed in the lens material. As for the impedance, and attending to focusing quality, any value can be optimal if we know exactly the properties of medium, transducer and backing. Since their exact values are difficult to know, it is as much as saying that not all impedances values are optimal, and the only robust value that does not require the exact knowledge of all agents, is an acoustic impedance equal to that of water.

Thus, this study is centered on the effect of the sound speed of the polymer, highly related to the thickness of the lens. Particularly this work proves that there is an optimum range of lens thicknesses, given that analytical models work best for fine lenses, but the discretization in voxels associated with 3D printing is a greater source of error the finer the lens. The set of numerical and experimental evaluations leaves a series of clear evidences. If we divide the materials into three groups of p-wave speeds, such as (i) notably lower than water (ii) near water, and (iii) notably higher than water, it becomes clear, both from theoretical analyses and simulations, that the second group should be avoided. On the other hand, attending to numerical simulations, optimal lens thicknesses are in the same range, between 10 and 25 voxels, for both slow and fast lenses, although fast lenses have certain advantages.

The main advantages associated with fast lenses are the following: (i) in fast lenses the presence of s-waves contaminates the main focus in a less risky way than in slow lenses, (ii) its behaviour is much more robust against eventual manufacturing errors that result in lenses with a p-wave speed different than expected, and (iii) rigid crystalline polymers employed for fast lenses have smaller acoustic absorptions than more deformable amorphous thermoplastics. Slow lenses, especially if they are made of elastomers in which the formation of waves can be neglected, have a higher energy efficiency associated with the value of their impedances, but do not offer clear advantages if the focusing quality is strictly evaluated.

Transcranial aberration correction based on 3D printed lenses was proposed last year as a technology capable of competing with the currently established solution based on phased arrays. The advantages of the 3D lenses are fundamentally related to the lower initial acquisition cost of the equipment, whereas the phased arrays can be straighter forward in their later use. However, apart from practical criteria of ease or cost, scientific criteria regarding the focusing accuracy must be considered when choosing one of these alternatives. In future works we will deal with the exhaustive comparison between focusing quality achieved by 3D lenses and phased arrays; so the present study was proposed as a necessary step prior to such a comparison between both technologies.

**Author Contributions:** Cconceptualization, JM.B., J.R. and M.F.; methodology, JV.S., N.J. and M.F.; software, J.R., N.J., and M.F.; validation, S.J. , N.J. and F.C.; formal analysis, JM.B, J.R, N.J, F.C and M.F.; writing, M.F.; review and editing, JV.S, S.J, N.J. and F.C.; supervision, JM.B and M.F.; project administration, JV.S. and F.C..; funding acquisition, JV.S. and F.C.

**Funding:** This work was partially supported by the Spanish "Ministerio de Economia y Competitividad" under the projects TEC2015-68076-R and TEC2016-80976-R. N.J and S.J. acknowledge financial support form Generalitat Valenciana through Grants No. APOSTD/2017/042, No. ACIF/2017/045, and No. GV/2018/11. F.C acknowledges financial support from Agència Valenciana de la Innovació through Grant No. INNCON00/18/9 and European Regional Development Fund (Grant No. IDIFEDER/2018/022).

**Acknowledgments:** The authors would like to thank José Sepúlveda, director of *Asociac nxión I2CV*, and MD. Xavier Serres for their important input and scientific support.

**Conflicts of Interest:** The authors declare no conflict of interest